\documentclass[a4paper]{spie}  
\usepackage[]{graphicx}
\usepackage{aas_macros}

\title{Dark tip-tilt sensing} 

\author{Arcidiacono Carmelo\supit{a,b},  Ragazzoni Roberto\supit{b,c}, Viotto Valentina\supit{b,c} , 
Bergomi Maria\supit{b,c}, Farinato Jacopo\supit{b,c}, Magrin Demetrio\supit{b,c}, Dima Marco\supit{b,c}, Gullieuszik Marco\supit{b,c}
and Marafatto Luca\supit{b,c}
\skiplinehalf
\supit{a} INAF - Osservatorio Astronomico di Bologna, Via Ranzani 1, I-40127 Bologna, Italy; \\
\supit{b} ADONI - Laboratorio Nazionale Ottiche Adattive, Italy;\\
\supit{c} INAF - Osservatorio Astronomico di Padova, Vicolo dell'Osservatorio 5, I-35141 Padova, Italy;
}

\authorinfo{Further author information: (Send correspondence to Carmelo Arcidiacono)\\C.A.: E-mail: carmelo.arcidiacono@oabo.inaf.it, Telephone:  +39 051 2095 704}

\begin{document} 
  \maketitle 

\begin{abstract}
Dark wavefront sensing in its simplest and more crude form is a quad-cell with a round spot of dark ink acting as occulting disk at the center. This sensor exhibits fainter limiting magnitude than a conventional quad-cell, providing that the size of the occulting disk is slightly smaller than the size of the spot and smaller than the residual jitter movement in closed loop. We present simulations focusing a generic Adaptive Optics system using Natural Guide Stars to provide the tip-tilt signal. We consider a jitter spectrum of the residual correction including amplitudes exceeding the dark disk size.
\end{abstract}


\keywords{Adaptive Optics, wavefront Sensing, Tip Tilt, Dark Wavefront Sensing}

\section{INTRODUCTION}
\label{sec:intro}  
The Tip-tilt sensing of a diffraction limited natural guide star subjected in closed loop to limited movements could represent the last obstacle of Laser Guide Star\cite{1985A&A...152L..29F} (LGS) based Adaptive Optics\cite{1953PASP...65..229B,Babcock253} (AO) systems on large and extremely large telescope\cite{2013JApA...34...81S,2007Msngr.127...11G,2012SPIE.8444E..1HJ}, basically setting up the ultimate performances of such systems. Peeling up its performances and reaching the as faint as possible magnitude limit become crucial into defining the performances in terms of achievable Strehl ratio and in sky coverage\cite{2014ExA....37..503A}.
While it is interesting that such a basic way of wavefront sensing finally would turn out to be of such paramount relevance, we revisit the already pointed out fact that most of the light involved in the tip-tilt measurement in a quad-cell approach is actually producing noise and not contributing to the accuracy or sensitivity of the measurement, see also Ragazzoni et al.\cite{RagazzoniDHO}. Specifically, under the class of “dark wavefront sensing”, tip-tilt measurements (or, better, “dark tip-tilt sensing”) aim to utilize just the smallest possible amount of photons contributing to the signal, while the vast majority being used for acquisition or to handle burst of jitter otherwise resulting into a loss of loop closure. The practical implementation of this approach does not involve any subtle diffraction limit effect, in contrast with similar realization for high order wavefront sensing, because the selection of the photons occurs in the exact plane where they a detected. 
We can envisage a CCD, conjugated to the focal plane, imaging the post-AO corrected Point Spread Function (PSF) of a reference star. Multiple dark spots of different size and placed at the crossroads of different groups of four adjacent pixels are present on the CCD, with the sensor locking onto different ones to achieve the ultimate performance, depending upon the residual jitter and energy concentration of the PSF.\\
We set up a bunch of realistic simulations using different jitter residuals and reference star intensities projecting onto a quad-cell the AO corrected image. We take into account different size of the dark disk in order to find out the optimal one. The dark tip-tilt control system is simulated by a pure integrator using a gain for each combination of jitter residual, intensities and disk radius. The applied gain value is the result of an optimization ending with the best one for each of the parameters combination considered.\\
We assumed three temporal series describing a jitter model characterized by two 2-dimensional oscillation frequencies with different amplitudes. Eventually we outline the comparison of the performances with respect to conventional quad-cell treated with the same numerical approach along with the first crude conclusions for such a kind of first order dark wavefront sensing.
\section{Dark Tip Tilt principle}
\label{sec:dtt}
\begin{figure}[h]
  \begin{center} 
		\includegraphics[height=6cm]{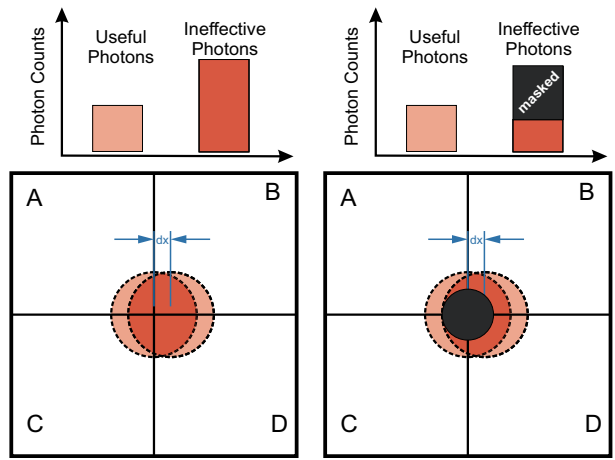}
	\end{center}
	\caption[]{\label{fig:sketch}
	On the left the usual quad-cell configuration. All the light in the four quadrants is measured and used for the signal reconstruction. But only the part corresponding to the non overlapping region before and after the image shift is actually contributing to the signal. On the right the same scenario, but in this case part of the ineffective photons are not considered since they are masked by a dark disk.}
	\end{figure}

The idea of coupling an occulting mask and a WFS has been proposed by Ragazzoni~(2004)\cite{2004SPIE.5382..456R} and later investigated in Le Roux and~Ragazzoni~(2005)\cite{2005MNRAS.359L..23L} generalized to the pyramid\cite{1996JMOp...43..289R} WFS case. Let's consider a simple quad-cell WFS. It's straightforward to demonstrate that given a reference source image larger than the jitter amplitude the WFS aims to measure, only a fraction of the photons composing the image contributes to the output tip-tilt signal. Referring to the sketch of a quad-cell conjugated to the focal plane in Figure~\ref{fig:sketch} and a misplacement on the horizontal axis, $dx$: only the faint colored photons at the two sides of the overlapping area of reference source images, before and after the misplacement, actually produce a valuable signal. On the other side the common part injures the signal in two ways: decreasing sensitivity and adding shot noise.
Tracing a dark area which occults the light in the ineffective region has the double advantage to increase sensitivity and reduce the photon noise. In order to obtain these advantages the dimension of the dark disk has to be smaller than the focal plane image of the reference source and in any case smaller than the typical misplacement experienced by the reference. \\ 
In the easiest realization the dark disk is actually a dark area on (or in front of) the CCD window. The dimension of the disk may vary being related to the atmospheric condition and on the effectiveness of the AO correction working to achieve smaller and smaller PSF size.\\ The masked photon may turn into a resource when the residual jitter exceeds the working range of the dark tip-tilt WFS or when the AO correction (or seeing) is getting worse. The AO designer may provide dark disks of different size, at specific positions of the detector (assumed it is something more than a 2$\times$2 pixel array) at the center of as many quadrants, in order to best match the actual residual wavefront the reference is experiencing.

\section{Simulations} 
 
We aim to investigate the closed loop behaviour of a Dark Tip-Tilt sensor (in the following only ``Dark Tip-Tilt") under AO corrected wavefront conditions. We may basically define the Dark Tip-Tilt by a few parameters. In our model we consider an ideal detector that images the PSF of a 8m diameter generic telescope. 
The PSF is modeled as an Airy function representing the diffraction limited PSF of the telescope. We choose a fixed and not optimized Field of View (FoV) for the sensor of the order of 35$\lambda$/D (D is the diameter of the telescope) at a working monochromatic wavelength of 0.5$\mu$m.
We realize the dark sensing concept by considering at the center of the Field of View, corresponding to center of the physical area of a square detector, a dark circular occulting mask. Herein we consider several radii for this mask of the order of half of the $\lambda$/D reference size.\\
\subsection{Jitter spectrum}
\begin{figure}[h]
  \begin{center} 
		\includegraphics[height=10cm]{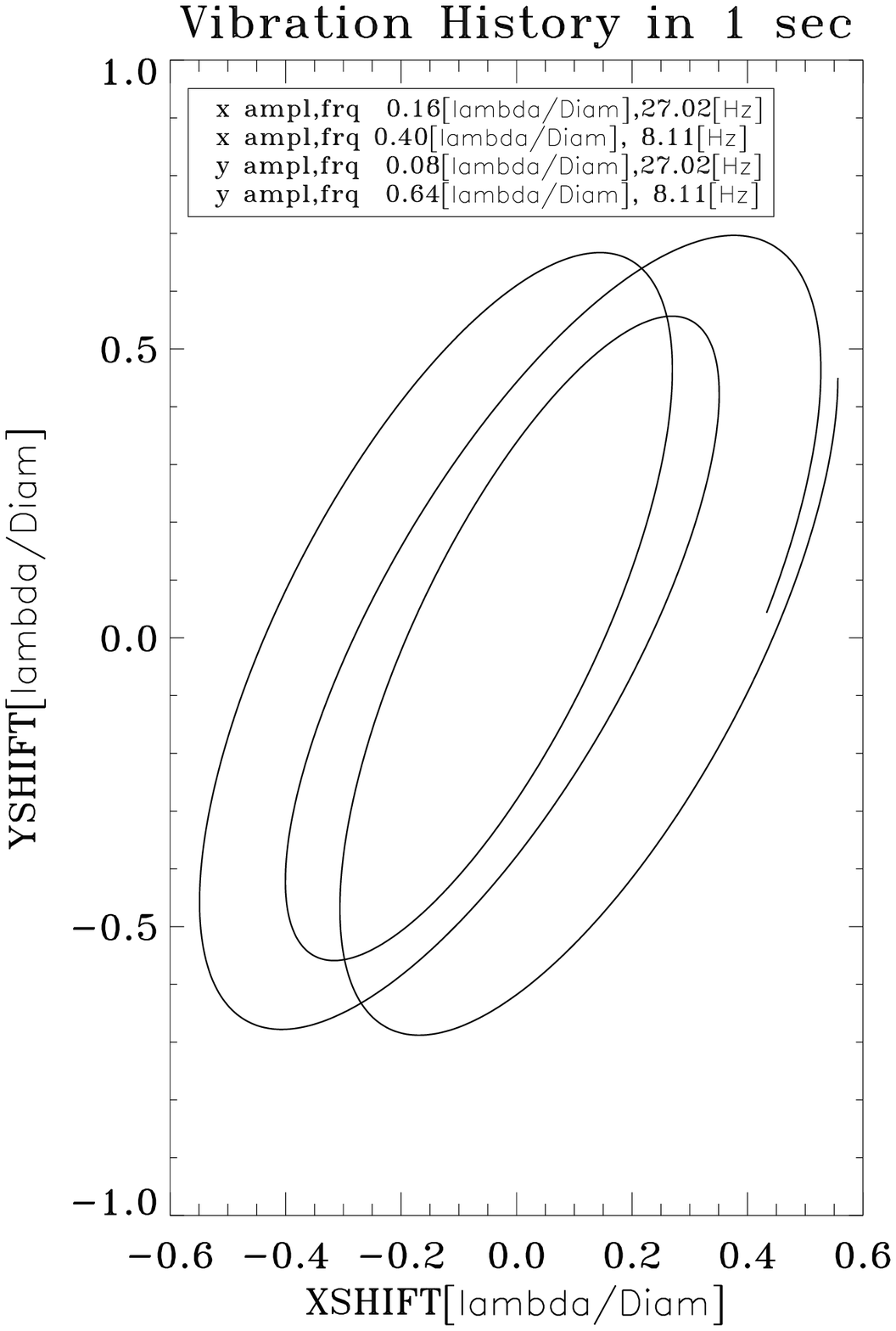}
    \includegraphics[height=10cm]{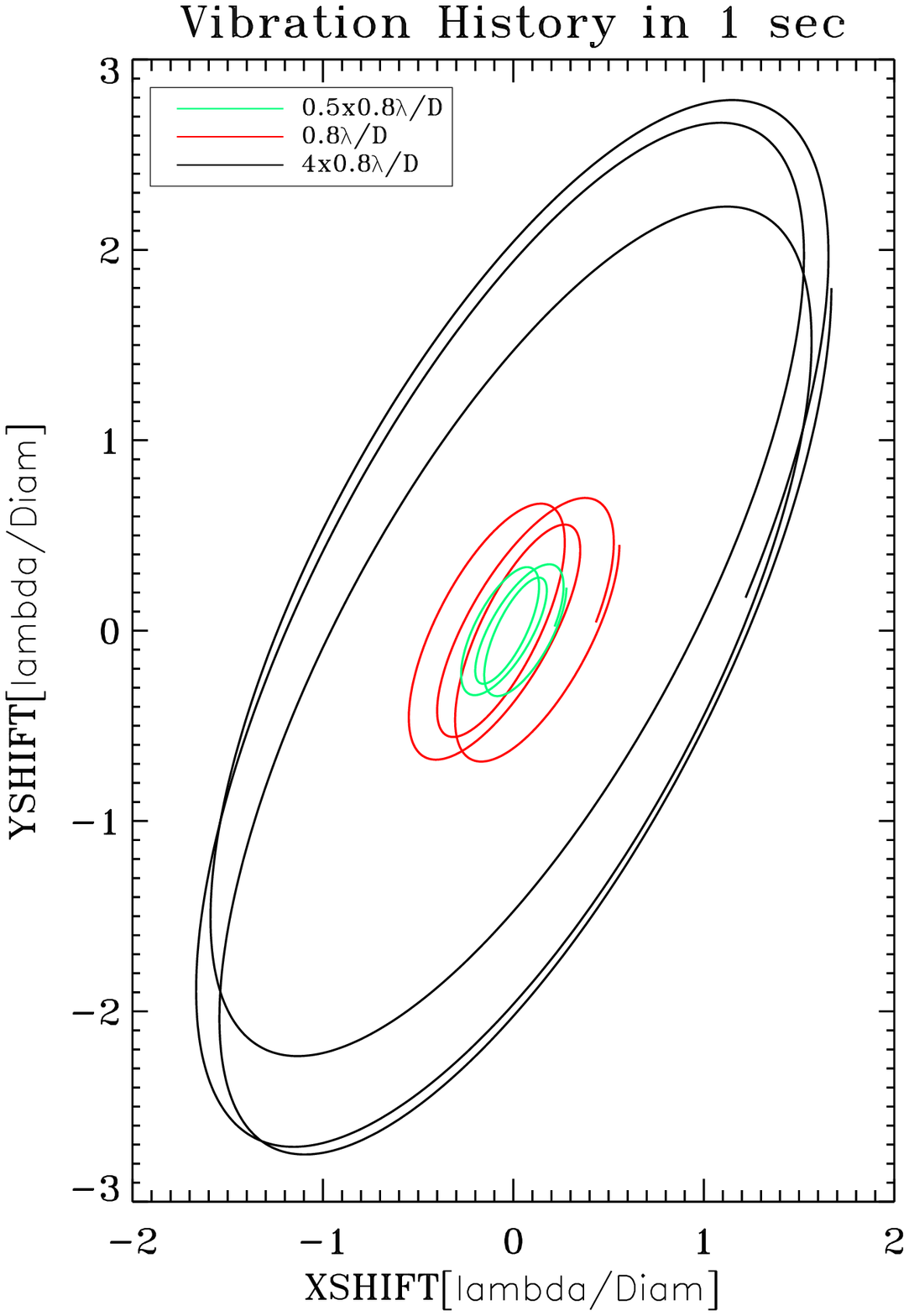}\\
	\end{center}
   \caption[]{ \label{fig:vibration} 
On the left the trajectory of the open loop PSF in terms of $\lambda$/Diameter for the reference case with amplitude 0.8$\lambda/D$. On the right the trajectories for all the three cases considered.	}
   \end{figure}

We are aware, because of geometrical considerations and because of the intrinsic small linear range, that the performance of the Dark Tip-Tilt are dependent on the residual jitter of the AO corrected PSF, higher the post-AO tip tilt corrected residuals, larger the amplitude of the oscillations. We considered three different jitter models, see Figure~\ref{fig:vibration}. The vibrations model is built upon two main realistic frequencies, $\nu$, at 8Hz and 27Hz examples of residual belonging to the typical range of frequencies envisaged on working AO systems\cite{2011aoel.confE...3E}.\\
Typically the residual vibrations of the images of an AO corrected camera are the results of servo lag error and of differential vibrations of the imaging camera with respect to the main AO wavefront sensor (WFS). The two frequencies considered here account for these two possible phenomena: considering three different couple of amplitudes we test the ability of the Dark Tip-Tilt to properly counter correct for the oscillation and to constrain its working range.\\
	 \\
\newline	
We arbitrary define as reference case the following amplitudes expressed in terms of wavelength/diameter units: $A=(A_x,A_y)$,  (0.16,0.08) 	and (0.4, 0.64) for the 8Hz and 27Hz respectively. Aside of the reference case we studied the performance for amplitudes that are 4$\times$ and $\frac{1}{2}\times$  multiples respectively. For sake of simplicity we will call these cases ``Ref. @0.8$\lambda/D$", ``4x" and ``0.5x". The 0.8$\lambda/D$ value is the squared sum of the amplitudes of the reference case, see Figure~\ref{fig:vibration}.

\subsection{Toy model}
\begin{figure}[h]
  \begin{center} 
		\includegraphics[width=10cm]{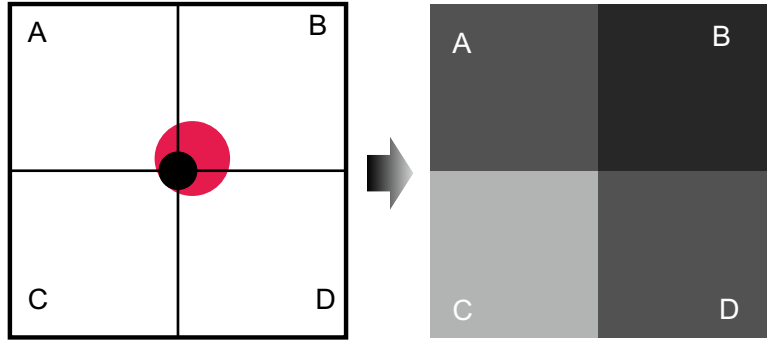}
		\end{center}
   \caption[example] 
   { \label{fig:quadcell} 
A graphical representation of the quad-cell. The capital letters A,B,C,D identify the quadrants and the light intensity measured therein. The reference star light is partially occulted by the dark disk at the center of the device. The arrow suggests that the numerical simulation transforms a 2D image into a quad-cell detection.}
   \end{figure}

We considered five different brightness intensities: 20, 50, 100, 500 and 1000 expressed in terms of photo-electron per frame. The specified intensities are the integral of the photo-electrons over the focal plane image as measured before the insertion of a field stop mask with a diameter of 35 $\lambda/D$. We set the exposure time of each frame to 1msec and a delay of the correction with respect to the measurement of 1msec too. The post AO PSF was designed as an Airy function corresponding to the diffraction limited PSF of an 8m telescope at the $0.5\mu m$ wavelength. However the results we obtained here may be easily scaled for different sensing wavelengths since all the geometric parameter are expressed in terms of $\lambda/D$ unit.
The Airy image moves because of the jitter with respect to the reference point. This point corresponds to the zero value of the slopes and also to the center of the disk drawn at center of the CCD. In the simulation the Airy function has a dimension of 512$\times$512 pixels and it is eventually binned to a 2$\times$2 to simulate the quad-cell. After the binning the photo-electron counts are processed to consider the Poisson photon (shot) noise. Please see for reference the Figure~\ref{fig:quadcell}\\
	
\begin{figure} [h]
  \begin{center} 
		\includegraphics[width=9cm]{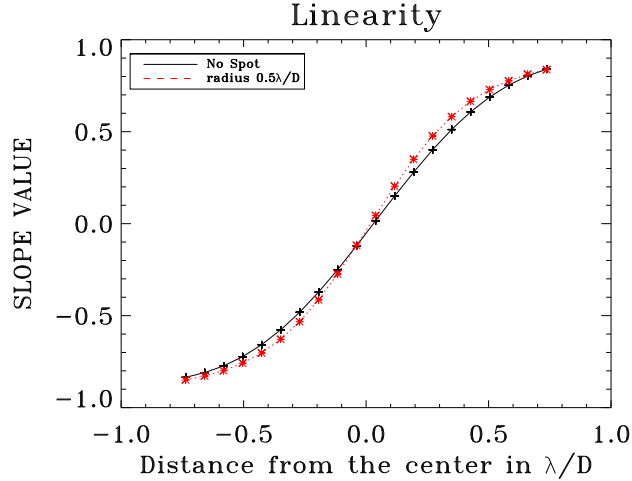}
		\end{center}
   \caption[example] 
   { \label{fig:linearity} 
In this plot the comparison of the quasi linear behaviour of the Dark Tip-Tilt as compared to the one of the unocculted quad-cell (dashed and solid line respectively). The dark disk for the occulted case has a radius of half of the sensing wavelength.}
   \end{figure}

From the quad-cell counts we get the couple of X and Y slopes ($S_x$,$S_y$) computed as:
\[
S_x = \frac{(A+C)-(B+D)}{A+B+C+D}  ; \\
S_y = \frac{(A+B)-(C+D)}{A+B+C+D}
\]
In order to realize the closure of the Tip-Tilt loop the slope vector ($S_x$,$S_y$) above is multiplied for the control matrix, without any correction for the non-linearity of the sensor for large slopes Figure~\ref{fig:linearity}. The control matrix is merely the pseudo-inverse of the Interaction Matrix (IM) of the system for the Tip and Tilt modes. We used the numerical model adopted for the simulation also for the purpose of the generation of the IM, considering the system working on a very bright source, with a negligible noise. The amplitude of both the tip and the tilt modes used for the registration of the IM was within the linear range of the WFS.\\

\section{Analysis}
\begin{figure}[h!]
   \begin{center}	
   \begin{tabular}{c}
	   \includegraphics[height=11cm]{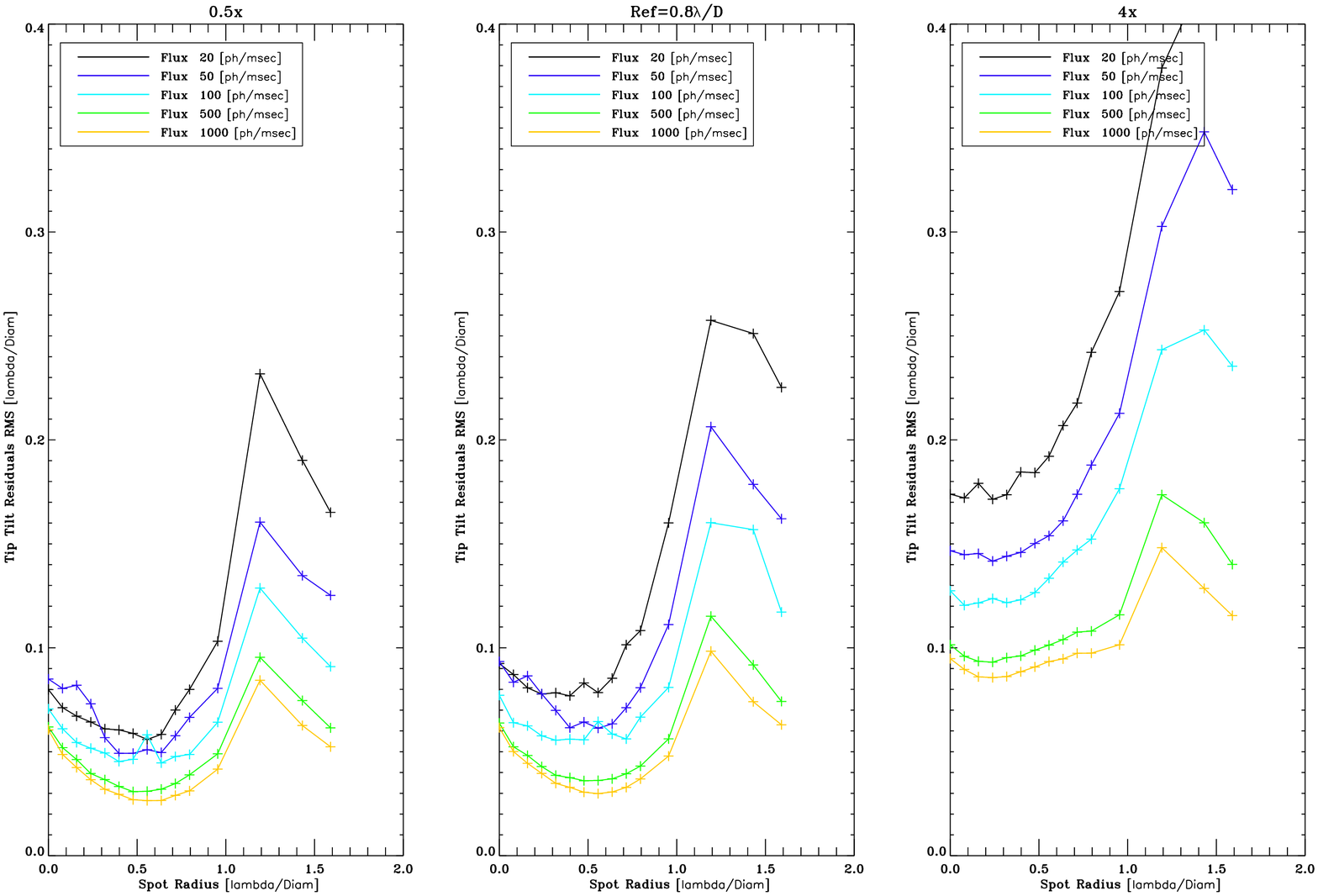}\\
   \end{tabular}
   \end{center}
   \caption[example] 
   { \label{fig:res} 
The plots above show the residual jitter measured after the correction done using Dark Tip-Tilt WFS with different dark disk radii. For reference, we add the first point on the left of each curve representing the classical quad-cell performance (disk radius = 0). Different colors are representing different reference brightness. The three plots from the left to the right have large input jitter amplitudes. As soon as the injected vibrations have an amplitude smaller or comparable to the dimension of the PSF the application of an occulting mask with radius somewhat smaller returns better results.}
   \end{figure}

\begin{figure} [h!] 
   \begin{center}
   \begin{tabular}{c}
   \includegraphics[height=11cm]{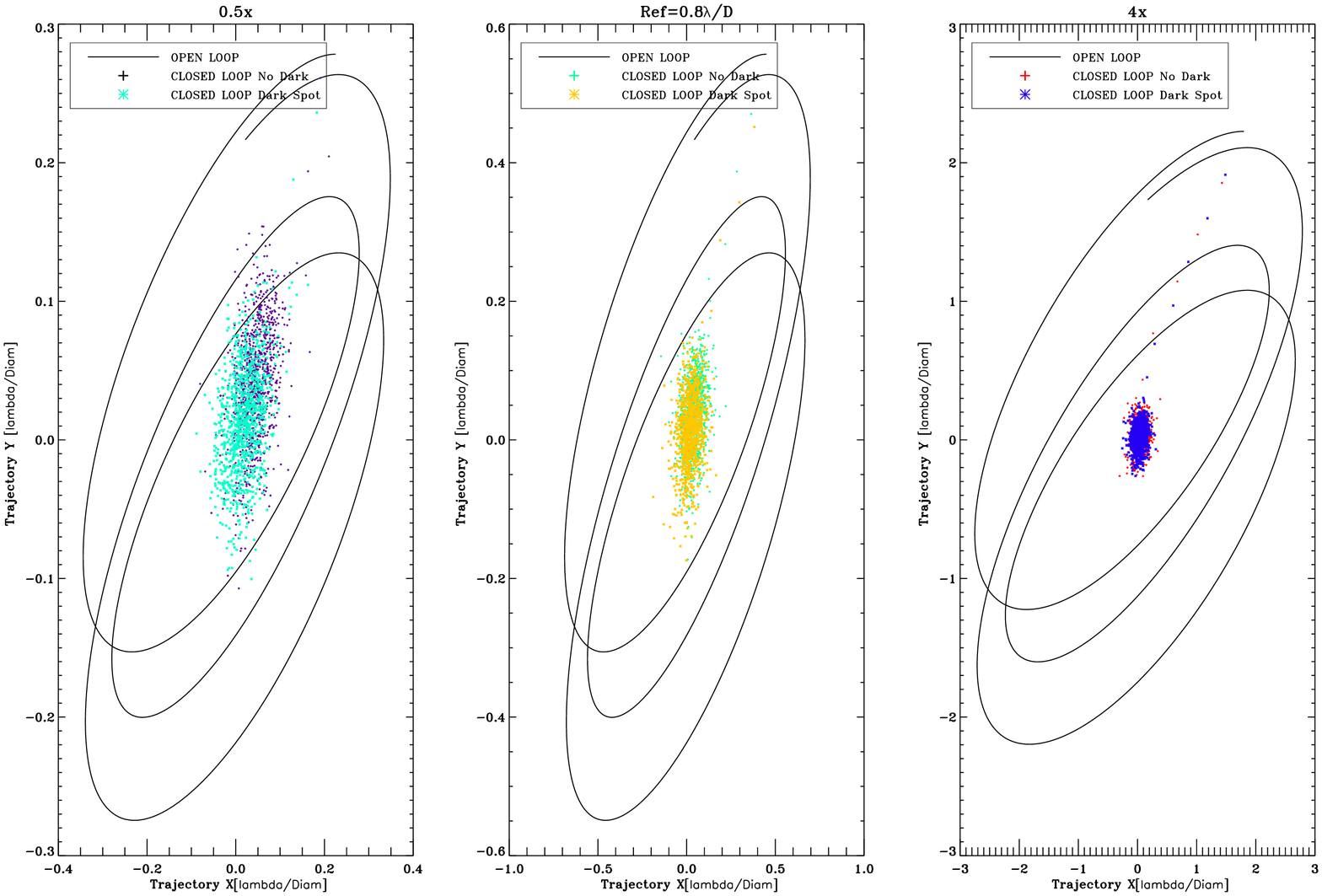}\\
   \end{tabular}
   \end{center}
   \caption[example] 
   { \label{fig:restrj} 
The superimposition of the open loop and closed loop trajectories for the three cases analyzed.}
   \end{figure}

The numerical simulations show, see Figure~\ref{fig:res}, that Dark Tip-Tilt may actually improve the quad-cell performance. A quick look
reveals what was easy to foresee: the performance in terms of vibration reduction depends on the radius of the occulting area, see Figures~\ref{fig:res} and~\ref{fig:restrj}. Actually
we find out that optimal values of the radius belong to .4-.7 $\lambda/D$ and such as radii return an advantage in terms of residual jitter with respect to the 
classical quad-cell for all the cases of flux and vibration amplitude explored.\\
In the high fluxes cases (500-1000 ph/msec) the gain in performance is evident, achieving a factor two gain in terms of final residual jitter. This result is in line with the results of the pyramid simulations in Le Roux and~Ragazzoni~(2005)\cite{2005MNRAS.359L..23L}.
	   \begin{figure}
   \begin{center}
   \begin{tabular}{c}
  \includegraphics[height=6cm]{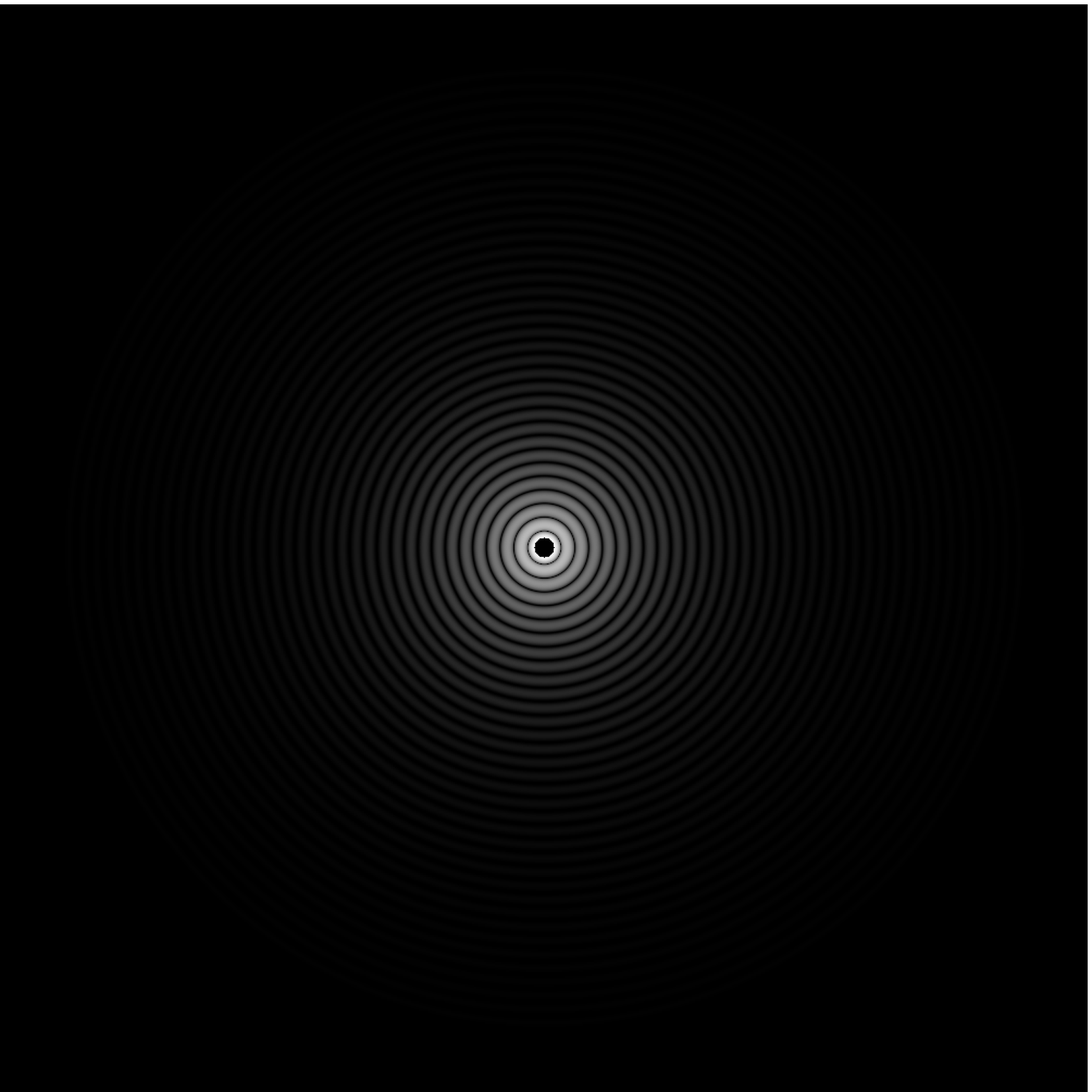}
	\includegraphics[height=6cm]{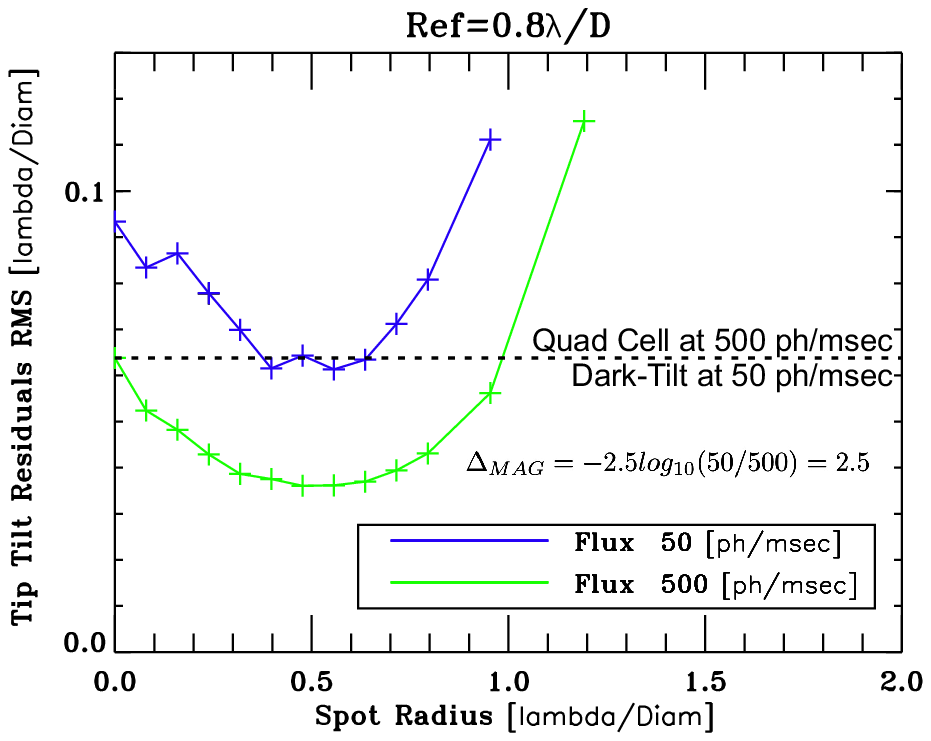}\\
	\end{tabular}
   \end{center}
   \caption[example] 
   { \label{fig:psf} 
On the left one of the best occulting masks, with radii 0.5 lambda / Diameter. On the right we show the gain in magnitude that may be achieved using the Dark Tip-Tilt sensor.}
   \end{figure} 

What we notice and we consider more intriguing is the behaviour of the proposed WFS with respect to the light intensity: the correction level achievable by the classical quad-cell at 500 counts has been obtained with only 50 counts per frame by the dark one (a factor 10 in flux or 2.5 magnitudes fainter) applying a mask with radius of $0.6\lambda/D$. See the reference case in Figure~\ref{fig:res}.\\
The analysis of the relation of the Tip-Tilt residuals versus the disk radius shows that the same level of jitter residuals may be obtained with a fainter star
with respect to the unmasked case for all the vibration amplitudes considered, see Figure \ref{fig:psf} .\\ 

\section{Discussion}
The numerical simulations we performed verify the expectations about the higher sensitivity of the Dark Tip-Tilt sensor with respect to 
the classical quad-cell. Looking into the magnitude of the reference star we found that the Dark Tip-Tilt gains up to 2~magnitudes .\\ The sensor achieve its optimal performances in a regime of jitter amplitude belonging more to a post AO corrected PSF than an open loop one. However it is also true that in our analysis the PSF was resembling the diffraction limited one being modeled as an Airy function. \\ Generally speaking the $\lambda/D$ of the analysis above may be substituted by the $\lambda/r_0$, where the $r_0$ is the open loop coherence length of the optical turbulence. With this substitution the Dark Tip Tilt sensor we have outlined in this study would be suitable also as main Tip-Tilt sensor of an adaptive optics system (for example in a Laser Guide Star system). As mentioned in section~\ref{sec:dtt} the AO designer may foresee different conditions of correction quality and provide a detector with multiple Dark Tip Tilt Sensor characterized by occulting masks with different sizes.\\
The actual performance in open loop or quasi open loop conditions may be affected by other characteristics, for instance the speckles number and size and chromatic effects. A more general study is needed to assess the behaviour of the Dark Tip Tilt sensor as main tip tilt sensor of an Adaptive Optics (or Multi Conjugate Adaptive Optics) system and such as analysis goes beyond the boundaries of this paper.

While this show that it does exists conditions under which the dark tip-tilt sensing is one order of magnitude more efficient than the conventional one, we would stress that we did not carry out any specific optimization of the dark disk size, so there could be conditions that would even more favor this kind of approach. Furthermore, as pointed out in Ragazzoni~(2015)\cite{AOELT4} it would be relatively easy to deposit masks of different sizes onto the cross of four pixels into a single CCD and switching with a service mirror from one to another to use the most effective dark disk size. Also, a zoom optics changing the actual scale could be used as well. It is noticeable that such an approach would also exhibit the potential benefits to adopting a field diaphragm that is variable accordingly to the disk size, eventually further optimizing by minimization of the background light.
\section{Conclusion}
We outlined the characteristics of a Dark Tip Tilt sensor: a quad-cell system opportunely masked to improve sensitivity and signal to noise ratio, ultimately improving the sky coverage.
We performed a numerical analysis showing that the system outperforms the classical quad-cell solution offering better residual or an increased limiting magnitude. However we notice that the actual gain will depend on the amplitude of the residual jitter the WFS as to afford: we obtained best sensitivity with amplitude somewhat smaller than the PSF size. \\ 
The analysis was considering a scenario in which the Dark Tip Tilt sensor was acting as a second level of correction on the back of a main adaptive optics loop. We also propose a possible extension of the method to open loop measurements: the extension of the working range to open loop conditions may ultimately improve the sky coverage of Laser Guide System\cite{2014SPIE.9148E..6FA} foreseen on the new class of extremely large telescopes in a very simple way. \\

\acknowledgments     
A special thanks to the LINC-NIRVANA\cite{2003SPIE.4839..536R,2012SPIE.8447E..0VC} team supporting the first author during the preparation of the manuscript.
\newpage
\bibliography{biblio}   
\bibliographystyle{spiebib}   

\end{document}